\documentclass[%
 reprint,
superscriptaddress,
 amsmath,amssymb,
 aps,
]{revtex4-2}

\usepackage[utf8]{inputenc}
\usepackage[T1]{fontenc}
\usepackage{amsmath}
\usepackage[normalem]{ulem}
\usepackage{dcolumn}% Align table columns on decimal point
\usepackage{bm}% bold math
\usepackage{makecell}
\usepackage{multirow}
\usepackage{xcolor}
\usepackage{graphicx}% Include figure files
\usepackage{dcolumn}% Align table columns on decimal point
\usepackage{bm}% bold math
\usepackage{gensymb}
\usepackage{siunitx}
\usepackage{booktabs}
\usepackage{subcaption}
\usepackage{ragged2e}
\usepackage{caption}
\begin{document}
\preprint{APS/123-QED}
\title{Interpretable machine learning to understand the performance of semi local density functionals for materials thermochemistry}
\author{Santosh Adhikari}
 \affiliation{Department of Chemistry and Biochemistry, University of South Carolina, Columbia, SC-29208, USA}
\author{Christopher J. Bartel}
 \affiliation{Department of Chemical Engineering and Materials Science, University of Minnesota, Minneapolis, MN 55455, USA}
\author{Christopher Sutton}
\email{cs113@mailbox.sc.edu}
\affiliation{Department of Chemistry and Biochemistry, University of South Carolina, Columbia, SC-29208, USA}
 
% \date{\today}

\begin{abstract}
This study investigates the use of machine learning (ML) to correct the enthalpy of formation ($\Delta$H$_f$) from two separate DFT functionals, PBE and SCAN, to the experimental $\Delta$H$_f$ across 1011 solid-state compounds. The ML model uses a set of 25 properties that characterize the electronic structure as calculated using PBE and SCAN. The ML model significantly decreases the error in PBE-calculated $\Delta$H$_f$ values from an mean absolute error (MAE) of 195 meV/atom to an MAE = 80 meV/atom when compared to the experiment. However, a similar reduction in the MAE was not observed for SCAN. Rather, the errors from the ML model (MAE = 76 meV/atom) and SCAN (MAE = 85 meV/atom) were observed to be comparable. To explain the substantial decrease in the error of PBE $\Delta$H$_f$ values and less so for SCAN, we employed partial dependence plots (PDPs) of an interpretable model, specifically generalized additive models (GAMs). The PDP+GAM approach allowed for an examination of the impact of all 25 features on the errors associated with the PBE and SCAN $\Delta$H$_f$ values. For PBE, the PDP+GAM analysis shows compounds with a high ionicity (\textit{I}), i.e., $I$>0.22, have errors in $\Delta$H$_f$ that are twice as large as compounds having \textit{I} < 0.22 (246 meV/atom compared to 113 meV/atom).  Conversely, no analogous trend is observed for SCAN-calculated $\Delta$H$_f$s, which explains why the ML model for PBE can more easily correct the systematic error in calculated $\Delta$H$_f$s for PBE but not for SCAN. Subgroup discovery (SGD) is used to better understand the relationship between the electronic structure features and the errors in the PBE calculated $\Delta$H$_f$. Out of these 25 features, SGD identifies the most reliable region or lowest error subgroup (108 meV/atom) to be comprised of 368 compounds (out of 1011 total) containing low charge transfer between atoms based on the selector \textit{I}<=0.52, $\zeta_{D}$<=0.43, and $P_D$<=0.02. Interestingly, although the literature suggests PBE is reliable for intermetallics but less so for oxides and halides, our analysis reveals intermetallics pose a challenge for PBE only when the charge transfer is significant (\textit{I}>0.22). Meanwhile, oxides and halides may be described accurately by PBE for systems in which charge transfer is relatively low (\textit{I} < 0.22). 
\end{abstract}

\maketitle

%\tableofcontents
\section{Introduction} 
Enthalpy of formation ($\Delta$H$_{f}$) and enthalpy of decomposition ($\Delta$H$_{d}$) are the two critical factors for understanding the stability of materials \cite{bartel2022review}. Because $\Delta$H$_{d}$ is calculated from the convex hull construction using $\Delta$H$_{f}$ for the set of relevant compositions, the development of effective computational approaches for materials design primarily relies on the accurate calculation of $\Delta$H$_{f}$. Density functional theory (DFT) approximations based on generalized gradient approximation (GGA) are widely applied to calculate $\Delta$H$_{f}$ for solid-state materials. Although the accuracy of DFT-calculated $\Delta$H$_{f}$ can depend on the choice of functional, PBE \cite{pbe1996} is the most common GGA that has been used to several hundred of thousand of compounds in open computational databases such as the Materials Project (MP) \cite{jain2011high,jain2013commentary}, the Open Quantum Materials Database (OQMD) \cite{saal2013materials,kirklin2015open}, and AFLOW \cite{curtarolo2012aflow}.  

However, the reliability of the calculated $\Delta$H$_f$ using PBE depends on the specific material class. For instance, PBE-calculated $\Delta$H$_{f}$ values are reasonably accurate for intermetallic alloys, such as Al-based and transition metal-based alloys \cite{wolverton2006first, isaacs2018performance}. On the other hand, PBE is unreliable in predicting $\Delta$H$_f$ for systems that combine metals and non-metals, such as oxides \cite{wang2006oxidation,hinuma2017comparison} and nitrides \cite{fuchs2002cohesive}. 

The primary source of these increased errors is related to the significant self-interaction error (SIE) inherent in semilocal functionals such as PBE \cite{perdew1981self,mori2008localization}. 
The meta GGA functionals such as SCAN \cite{scan2015}, have been demonstrated to enhance the accuracy of $\Delta$H$_{f}$ predictions by as much as  50\%  relative to PBE \cite{isaacs2018performance}. However, issues remain in overestimating $\Delta$H$_{f}$ for compounds that are considered to be \textit{weakly bound} (typically indicated by $\Delta$H$_{f}$ values between -1.0 eV/atom to -0.5 eV/atom) \cite{isaacs2018performance}, such as in the case of intermetallics. A recent advancement of SCAN, known as r$^2$SCAN \cite{r2scan}, has been shown to further improve the MAE of SCAN by approximately 15 meV/atom for a dataset of over 1000 solids \cite{kingsbury2022performance}. However, the persistent problem of overestimation for intermetallics still remains, as pointed out in Ref \cite{kothakonda2022testing}. Alternatively, methods like random phase approximations (RPA) and hybrid functionals, such as HSE06 \cite{krukau2006influence}, address some of the limitations of GGA functionals by, for example, incorporating nonlocality. However, these methods are computationally demanding and may not always lead to higher accuracies in the predicted $\Delta$H$_{f}$ values \cite{ong2011comparison,nepal2020formation}. 

A cheaper approach to improving PBE-calculated $\Delta$H$_{f}$ ($\Delta$H$_{f}^{PBE}$), which is widely used in databases such as MP, OQMD, and AFLOW, is to apply an on-site Hubbard U to $d$- or $f$-orbitals to reduce the effect of SIE \cite{dudarev1998electron,wang2006oxidation}. However, determining the appropriate +U value remains an open question, and different strategies have been proposed, such as performing self-consistent calculations using linear response \cite{cococcioni2005linear,kulik2006density,shishkin2016self} and tuning +U to recover higher accuracies for specific properties \cite{loschen2007first,singh2015putting}. Additionally, since +U corrections are commonly applied only to states with $d$ or $f$ characters (eg. strongly correlated materials), SIE associated to states with $s$ and $p$ characters may still remain \cite{bajaj2017communication,moore2022high}. 

It is also possible to correct much of the errors made by GGA (or GGA+U) functionals by fitting corrections to the elemental reference energies. This approach was first shown for oxides \cite{wang2006oxidation} and generalized to compounds spanning the periodic table  \cite{lany2008semiconductor,stevanovic2012correcting,jain2011high,kirklin2015open}. Fitted corrections are most effective when errors in $\Delta$H$_{f}$ are systematic with respect to certain elements (GGAs) and were shown to yield less of an improvement for SCAN than for PBE \cite{bartel2019role}.

Alternatively, machine learning (ML) has conventionally been applied to predict the DFT-calculated $\Delta$H$_{f}$ of solids at a large scale (> 1000 materials) \cite{ward2016general,deml2016predicting,zhang2020machine,meredig2014combinatorial,elemnet,goodall2020predicting,huang2020practicing,schnet,cgcnn,megnet}. However, these models inherit the underlying errors of the DFT functional (typically PBE). A more recent study \cite{gong2022calibrating} instead uses DFT-calculated $\Delta$H$_{f}$s reported in the MP database (computed using PBE with elemental corrections) to predict experimental $\Delta$H$_{f}$s ($\Delta$H$_{f}^{expt}$) with 30\% more accuracy compared to the MP database using compositional \cite{ward2016general} features (i.e., just considering the chemical formulas of each compound). Although this study reports a substantial correction to the DFT $\Delta$H$_{f}$ errors, it does not explain the evolution or origin of the error. An understanding of $\Delta$H$_{f}$ errors due to the selection of a specific DFT functional, necessitates using properties as features beyond compositional ones. These properties should be able to characterize the electron density distribution computed by the chosen functional.
 
In this work, we use ML to both understand where the DFT-calculated $\Delta$H$_{f}$ values are reliable and correct them to align with the corresponding corresponding $\Delta$H$_{f}^{expt}$ values. To achieve this, we initially formulate a set of numerical electronic-structure-based features that are sensitive to the choice of functional, for example, charge transfer and the density of states. We subsequently analyze the influence of these features using an interpretable ML approach to explain the varying degrees of improvement in 
$\Delta$H$_{f}^{DFT}$ for two DFT functionals, PBE and SCAN.

\section*{Dataset and features}  
 This work utilizes the dataset of $\Delta$H$_{f}^{expt}$s previously examined by Bartel \textit{et al.} \cite{bartel2019role}, which consists of 714 binary, 270 ternary, and 28 quaternary compounds. The dataset spans over 62 elements across the periodic table and a diverse set of chemical families such as oxides, sulfides, nitrides, phosphides, halides, and intermetallics. \\

To generate the electronic features, we use LOBSTER (version 4.1.0) \cite{maintz2016lobster} and Density Derived Electrostatic and Chemical Methods (DDEC6, version 3.5) \cite{manz2016introducing,limas2016introducing,manz2017introducing}, which post-processes the outputs from VASP calculations (see METHODS section) to generate a set of features (described below in more detail). The initial dataset by Bartel \textit{et al.} \cite{bartel2019role} contained 1012 compounds; however, we were only able to generate all the features summarized in Table \ref{input features} for 1011 compounds using PBE and 984 compounds using SCAN because of issues with post-processing via LOBSTER and DDEC6. 
 \begin{table*}[htb]

    \caption{Features used as inputs into our ML model}
    \centering
    \resizebox{2.0\columnwidth}{!}{
    \begin{tabular}{cccc}
    \toprule
        Features & Label & Definition & Unit \\
        \midrule
        \multirow{6}{*}{\makecell{LOBSTER}} & \makecell{\textit{$\mu$$_{HP}$, MAX$_{HP}$},$~\sigma${$_{HP}$}} & \makecell{averaged, maximum, and standard deviation value of the integral of pCOHP} & eV \\
        & \makecell{\textit{$\mu$$_{OP}$, MAX$_{OP}$},$~\sigma${$_{OP}$}} & \makecell{averaged, maximum, and standard deviation value of the integral of pCOOP} & eV \\ &
        \makecell{\textit{$\mu$$_{BI}$, MAX$_{BI}$},$~\sigma${$_{BI}$}} & \makecell{averaged, maximum, and standard deviation value of the integral of COBI} & eV \\
        &
        \makecell{\textit{s, p}} & \makecell{  relative contribution of s-, p-, d-, and f-orbitals} & \\
        &
         \makecell{\textit{d, f} } & \makecell{ within $\pm$ 3 eV relative to Fermi level} & -
        \\~\\
        \midrule
        \multirow{5}{*}{Structure} & \textit{N} & \makecell{number of atoms per unit volume of the unit cell} & at./{\AA}$^{-3}$ \\
        & \makecell{\textit{Eg}} & \makecell{band gap computed as the minimum difference of kohn-sham eigenvalues} & eV \\
        & \makecell{\textit{CN}} &  \makecell{average coordination number of atoms in the cell using Brunners Algorithm \cite{brunner1977definition}} & - \\
        & \makecell{\textit{$\rho$}} &  \makecell{density of the compound} & gm/cm$^3$ \\
        & \makecell{$\eta$} &  \makecell{product of $N$ and $CN$} & at./{\AA}$^{-3}$ \\~\\
        \midrule 
        \multirow{4}{*}{\makecell{DDEC6 and Bader \\ based-features}} & \makecell{\textit{$\zeta$$_{B}$}, \textit{$\zeta$$_{D}$} } & \makecell{average charge transfer using Bader and DDEC6 analysis} & e$^-$/at. \\
        & \makecell{\textit{P$_{B}$}, \textit{P$_{D}$}} & \makecell{dipole moment per unit volume from Bader and DDEC6 analysis}  & e$^-$/{\AA}$^2$ \\  
        & \makecell{\textit{p}} & \makecell{net dipole moment per unit cell from DDEC6 analysis}  & e$^-$/{\AA}$^2$ \\
        & \makecell{\textit{I}} & \makecell{ratio of net charge transfer to the summed bond order \cite{sun2019map}}  & e$^-$/at.
        \\~\\
        \midrule 
        \multirow{1}{*}{\makecell{Atomic}} & \makecell{\textit{Z}} &
        \makecell{a compound index based on modified Pettifor values \cite{glawe2016optimal} \\defined for constituent elements\\} & -- \\~\\
        \bottomrule
    \end{tabular}}
    \label{input features}
\end{table*}

 \subsection*{Structure-based features:}
Table \ref{input features} lists several structure-based features that are generated from the PBE and SCAN optimized geometries from Ref. \cite{bartel2019role}. These features includes the total number of atoms per unit volume ($N$) and the average coordination number ($CN$), which is defined as the nearest neighbors for each atom within a cutoff radius of 8 {\AA}. The $CN$ is calculated using the Brunners algorithm \cite{brunner1977definition} within pymatgen \cite{Ong_Python_Materials_Genomics_2013}. We have also included the product of \textit{CN} and \textit{N}, which is labeled as packing ($\eta$).

\subsection*{LOBSTER-based features:}
 LOBSTER \cite{maintz2016lobster} was used to calculate the integrated projected crystal orbital overlap population (IpCOOP) \cite{hughbanks1983chains}, projected crystal orbital Hamilton population (IpCOHP) \cite{dronskowski1993crystal} and crystal orbital bond index (ICOBI) \cite{muller2021crystal} for all atom pairs separated up to 5 {\AA}. IpCOOP, IpCOHP, and, ICOBI quantify the number of electrons, contribution to the band-structure energy, and bond index (degree of covalency, ionicity), respectively, associated with the given bond. As a result, a set of IpCOOP, IpCOHP, and ICOBI values are calculated for all unique pairs of atoms within the cutoff radius of 5 {\AA}. For example, for the ternary compound ABC, each of the three quantities will be calculated for all A-B, B-C, and A-C atom pairs within 5 {\AA}. To ensure that a consistent set of nine numbers are computed for each compound, the maximum, average, and the standard deviation were calculated across each unique pair of atoms for each compound (e.g., maximum, average, and standard deviation across the values generated for A-B, B-C, and A-C). If multiple IpCOOP, IpCOHP, and ICOBI values were produced for each unique pair of atoms in a given compound, potentially due to minor variations in the bond distance within the structure, we consolidated this information by taking the average of these values for each pair.
 
 The normalized contribution of \textit{s-}, \textit{p-}, \textit{d-} and \textit{f-} orbitals to the bands within an energy window of $\pm$3 eV around the Fermi level for each compound was calculated using LOBSTER.

\subsection*{DDEC6 features:}
Several features have been previously used to quantify the charge-transfer character in compounds such as the net dipole moment per unit cell ($p$) \cite{manz2016introducing,limas2016introducing} and ionicity ($I$) \cite{sun2019map} by evaluating the contribution of each atom in the system to the total dipole moment and the extent of ionic bonding, respectively. For illustration, $I$ for a quaternary compound A$_{\alpha}$B$_{\beta}$C$_{\theta}$D$_{\gamma}$ is computed as:
\begin{equation}
    I = \frac{1}{\alpha + \beta + \theta + \gamma} ({\alpha \frac{\delta_{A}}{s_A} + \beta \frac{\delta_{B}}{s_B} + \theta \frac{\delta_{C}}{s_C} + \gamma \frac{\delta_{D}}{s_D}})
\end{equation} where $\delta_{A}$ and $s_{A}$, respectively denote the net charges assigned and summed bond orders for the element A in the compound \cite{sun2019map} and so on. Building off of this previous work, the dipole moment per unit volume ($P$) was calculated according to the equation:  
\begin{equation}
    P = \frac{1}{2} \frac{l}{V} {\sum_{i=1}^{N} |Q_{i}|},
\end{equation}
where, $N$ is the total number of atoms, Q$_{i}$ is the net charges acquired (positive) or lost (negative) by the i$^{th}$ atom in the system, and $l$ is the distance between the center of positive and negative charges. 
The average charge transfer ($\zeta$) in the system is also calculated based on the equation: 
    \begin{equation}
        \zeta=\sum_{i=1}^{N}\frac{| Q_{i} |}{2N},
    \end{equation}
 In the expression of both $P$ and $\zeta$, the factor of 2 in the denominator compensates for the double-counting of the charge transfer (charge acquired vs. charge lost) and for all samples, both $P$ and $\zeta$ are greater than zero.\\
The net charges and dipole moment contribution of each atom computed with DDEC6 \cite{manz2016introducing,limas2016introducing,manz2017introducing}  was used to compute $p$ \cite{manz2016introducing,limas2016introducing}, $\it{I}$ \cite{sun2019map}, \textit{$\zeta$$_D$} and \textit{P$_D$}, where the subscript, $D$, indicates computation using the DDEC6 methods. Separate from DDEC6, we also computed the net charges on each atoms using Bader analysis \cite{henkelman2006fast} and calculated \textit{$\zeta$$_B$} and \textit{P$_B$} starting from these charges (hence, the subscript, $B$, for Bader). Both features are incorporated as the calculation of atomic charges differs when using DDEC6 in comparison to using Bader atomic charges, which are both widely used. Bader charges are calculated by dividing the electron density at zero flux surfaces, whereas DDEC6-based charges are calculated as a functional of the electron density. More importantly the computed net atomic charges in both cases are derived from the total electron density, which naturally incorporates the effects of SIE. 

\subsection*{Atomic features:}
In addition to these electronic features, the compositionally averaged modified Pettifor index \cite{glawe2016optimal} ($Z$) was also used to distinguish between compounds based on the chemical elements. Modified Pettifor's index is a unique value assigned to each element in the periodic table that gives the measure of the extent of its replaceability in the crystal structure.

 \section*{Methods}
DFT calculations using the PBE \cite{pbe1996} and SCAN \cite{scan2015} functionals were performed using the projector augmented wave (PAW) formalism as implemented in the Vienna ab initio simulation package (VASP) code version 5.4.4. The structure files required for the calculations were obtained from Ref \cite{bartel2019role}. We used the PAW pseudopotentials as recommended in the VASP manual, a plane-wave cutoff of 520 eV, energy convergence criterion of 10$^{-6}$ eV, the smearing parameter ($k_{B}T$) of 0.01 eV (following the first order Methfessel-Paxton scheme), and gamma-centered Monkhorst-Pack k-point grid with 20|b$_i$| discretizations along each
reciprocal lattice vector, b$_i$, for the Brillouin zone sampling in all the calculations.

The ML task here adopted the so-called ``$\Delta$'' learning scheme \cite{ramakrishnan2015big} where the electronic structure properties from DFT calculations (Table \ref{input features}) are used as model inputs to predict the difference between $\Delta$H$_{f}^{expt}$ and $\Delta$H$_{f}^{DFT}$, where the $DFT$ functionals are PBE and SCAN. This difference is then added back to $\Delta$H$_{f}^{DFT}$ to predict $\Delta$H$_{f}^{expt}$ (with $\Delta$-learned predictions denoted $\Delta$H$_{f}^{ML}$). Linear ridge regression (Linear), random forest regressor (RFR), and kernel ridge regression (KRR) using the Laplacian kernel (KRR+Lap) and the Gaussian rbf kernel (KRR+rbf) were performed with scikit-learn \cite{scikit-learn} version 1.1.1. All hyperparameters were tuned via grid search using $5$-fold cross-validation. The GAM models were trained using the pyGAM package \cite{pygam}. A linear response function (LinearGAM) was used in training the $\Delta$-learning model. We note that to evaluate the variability of each of the ML models, 51 distinct models were trained using 51 unique random seeds, leading to different random  80 \%  / 20 \% training/test splits for each model (see Table \ref{perf} for the mean accuracy and standard deviation of the test error). $\Delta$H$_{f}^{DFT/ML}$s were utilized to calculate the enthalpy of decomposition ($\Delta$H$_{d}^{DFT/ML}$) for stability analysis using the approach described in Ref \cite{bartel2020critical}.  
 
For consistency of the notation, we use ${\delta}\Delta$H$_{f}^{DFT}$ and ${\delta}\Delta$H$_{f}^{ML}$ to represent the difference between $\Delta$H$_{f}$ for the DFT functional and ML model, respectively, with experiment (i.e., ${\delta}\Delta$H$_{f}^{DFT/ML}~$=$~\Delta$H$_{f}^{expt}~$ - $~\Delta$H$_{f}^{DFT/ML}$). To evaluate model performance, the mean error (ME) and mean absolute error (MAE) of $\Delta$H$_{f}^{DFT/ML}$ are computed relative to $\Delta$H$_{f}^{expt}$. The ME (MAE) is computed by averaging ${\delta}\Delta$H$_{f}^{DFT}$ (|${\delta}\Delta$H$_{f}^{DFT}$|). Figures S1 - S25 display the distribution of ${\delta}\Delta$H$_{f}^{DFT}$s versus all 25 features listed in Table \ref{input features}. 
 
Finally, for our analysis of the errors for the functional with experiment, the partial dependence plots (PDPs), generated from the gamma response function (GammaGAM) were used to predict |${\delta}\Delta$H$_{f}^{DFT}$|.
 Both LinearGAM and GammaGAM are referred to as `GAM' in the results section. To specifically obtain regions where a given set of $\Delta$H$_{f}^{DFT}$ has a decreased error relative to experiment (i.e., reliable regions), we used subgroup discovery (SGD) as described in Ref. \cite{sutton2020identifying}. The SGD target variable for the identification of the reliable regions was the absolute  errors |${\delta}\Delta$H$_{f}^{DFT}$|. All computations were performed using the SGD implementation in realKD 0.7.2. 
 
\section*{Results and Discussion}
\subsection*{$\Delta$-learning $\Delta$H$_f$s}
Table \ref{perf} summarizes the average $\Delta$H$_f$ errors calculated from several ML methods.
\begin{table}[h!]
\centering
\setlength{\tabcolsep}{12.0 pt}
\caption{ Mean absolute errors (meV/atom) predicted by different machine learning (ML) models trained using the features from PBE  and SCAN listed in Table \ref{input features} for the 20\% left-out set. The MAEs for each ML model correspond to the median model across 51 runs. The standard deviations for the 51 runs (train/test splits) are also provided alongside. }

\resizebox{\columnwidth}{!}{%
\begin{tabular}{lccc}
\toprule
&  &  \multicolumn{2}{c}{MAE (meV/atom)}\\
\cmidrule(lr){3-4}
& ML models &  PBE features & SCAN features\\
\midrule
      & Linear            & 102 $\pm$ 6     & 89 $\pm$ 5     \\
                                 & GAM             & 91 $\pm$ 6      & 86 $\pm$ 7     \\
                                & KRR+rbf        & 92 $\pm$ 6      & 82 $\pm$ 5     \\
                                & KRR+Lap  & 84 $\pm$ 6      & 81 $\pm$ 5    \\
                                & RFR             & 80 $\pm$ 5      & 76 $\pm$ 5     \\
                                \bottomrule
\end{tabular}}
\label{perf}
\end{table}

\begin{figure}
%\centering
    \captionsetup[subfigure]{labelformat=empty}   
    \begin{subfigure}[b]{0.495\textwidth}
 \flushleft{(a)}
                \includegraphics[scale=0.39]{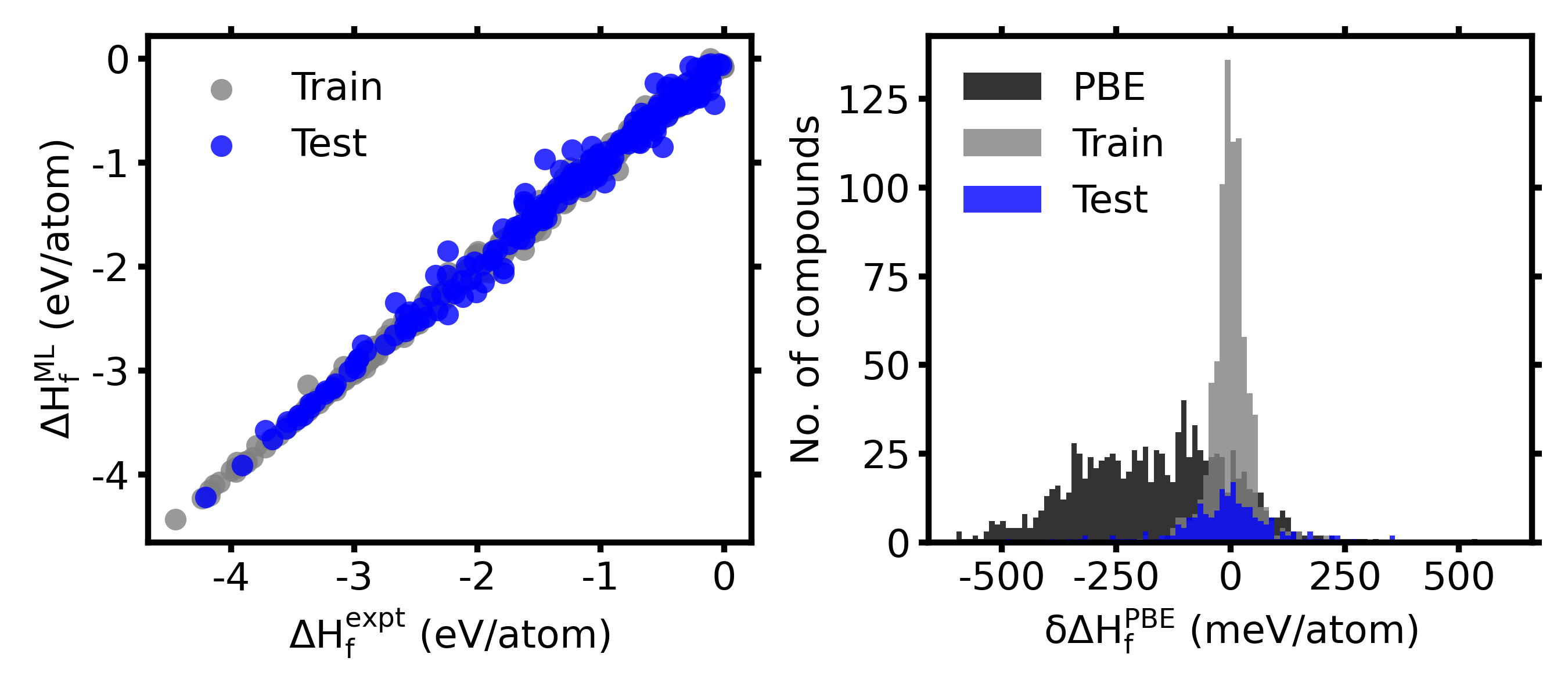}
    \end{subfigure}
        \begin{subfigure}[b]{0.495\textwidth}
  \flushleft{(b)}
                \includegraphics[scale=0.39]{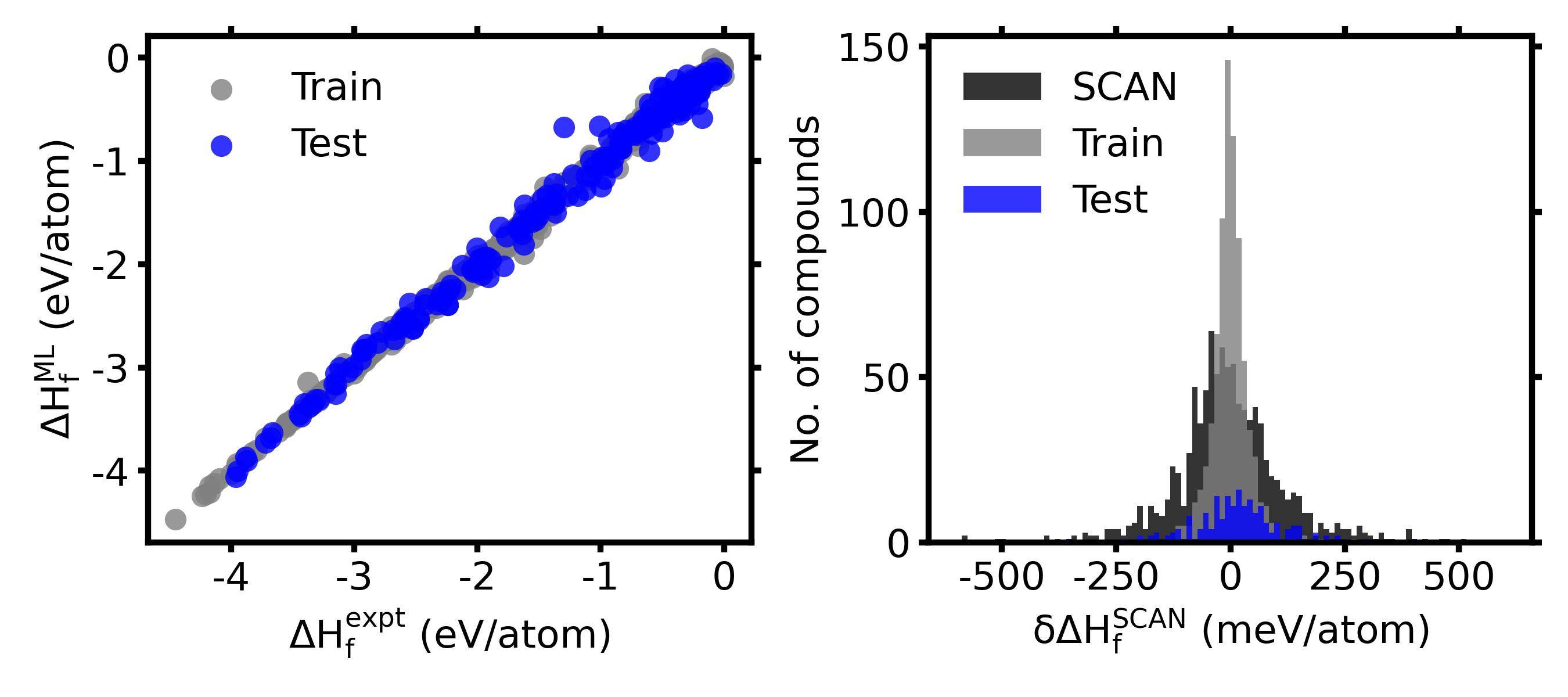}
    \end{subfigure}
    \caption{Parity plot of $\Delta$H$_{f}^{expt}$ and $\Delta$H$_{f}^{ML}$ and corresponding histograms of ${\delta}\Delta$H$_{f}^{DFT}$ (black) and ${\delta}\Delta$H$_{f}^{ML}$ (gray; for train-set and blue for test-set) for the RFR model based on (a) PBE and (b) SCAN (Table \ref{perf}). The mean absolute error, MAE, of ${\delta}\Delta$H$_{f}^{ML}$ is 80 meV/atom, and 76 meV/atom when used with PBE, and SCAN, respectively.}
    \label{lap:pbe}

\end{figure}
The highest performing $\Delta$-learned ML model for PBE was RFR, which has an MAE of 80 meV/atom for the 20\% test set which is substantially smaller than PBE (MAE = 195 meV/atom) for the same 20\% test set. Fig. \ref{lap:pbe} shows the distribution of ${\delta}\Delta$H$_{f}^{ML}$s, which is equally distributed above and below zero indicating that the ML model has a mean near zero and hence no systematic bias which is in stark contrast to PBE. The significant reduction in ${\delta}\Delta$H$_{f}^{ML}$ is attributed to the compounds with a large ${\delta}\Delta$H$_{f}^{PBE}$ (Fig. \ref{lap:pbe}), which trends in the number of elements in each compounds. More specifically, the set of compounds for which |${\delta}\Delta$H$_{f}^{PBE}$| > 200 meV includes 70\% of the ternary compounds (184 out of 270) and all the 28 quaternary compounds. The gain in accuracies with $\Delta$H$_{f}^{ML}$ compared with $\Delta$H$_{f}^{PBE}$ is factors of 7.5 ( 43 meV/atom vs. 322 meV/atom) for quaternary and 3.7 (69 meV/atom vs. 252 meV/atom) for ternary both of which are much higher than a factor of 2 for binary (86 meV/atom vs. 168 meV/atom).\par

\begin{figure}
\centering
    \captionsetup[subfigure]{labelformat=empty}
    \begin{subfigure}[b]{0.49\textwidth}
                \flushleft{(a)}
                \includegraphics[scale=0.53]{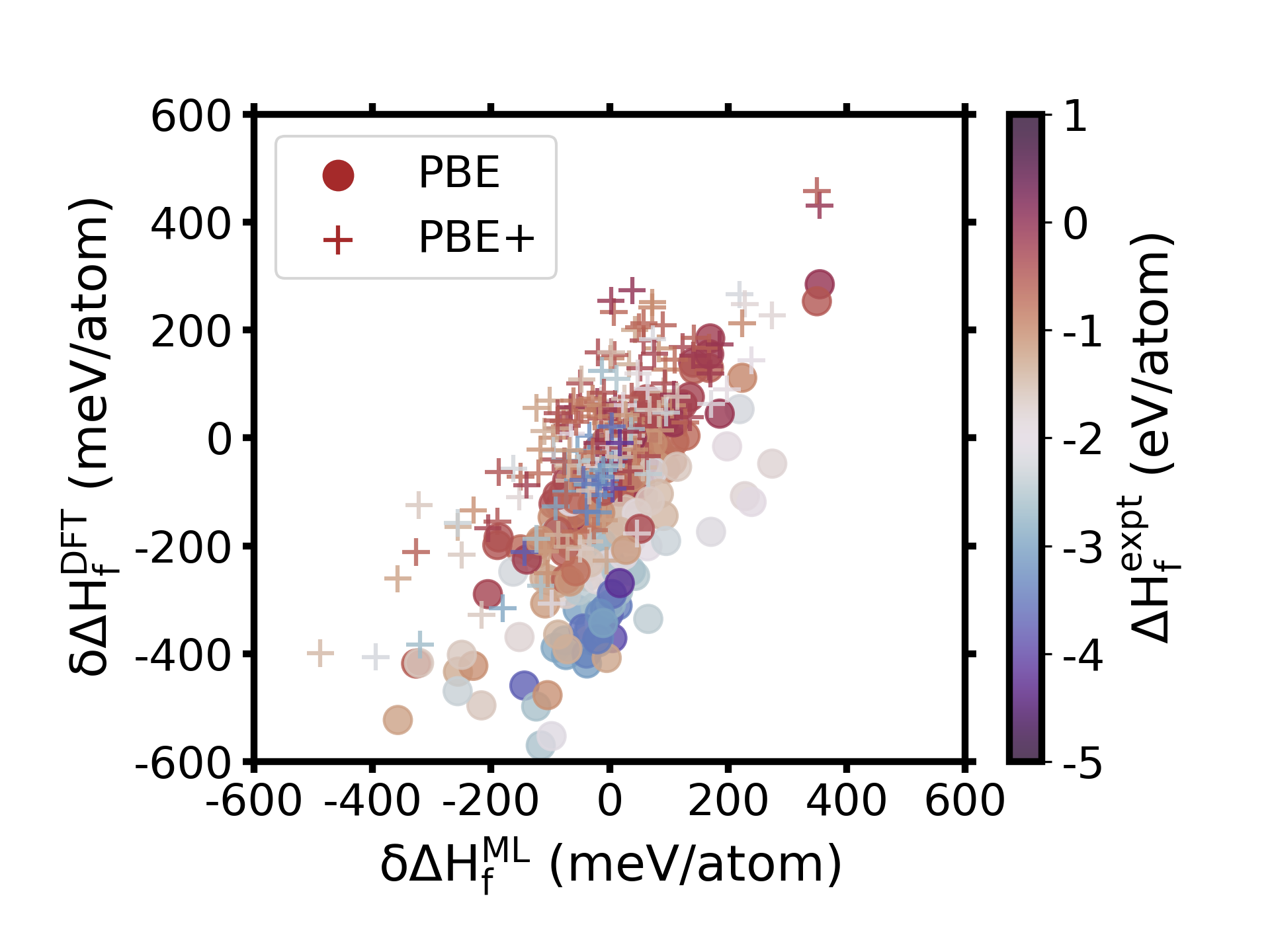}
    \end{subfigure}
        \begin{subfigure}[b]{0.49\textwidth}
  \flushleft{(b)}
                \includegraphics[scale=0.53]{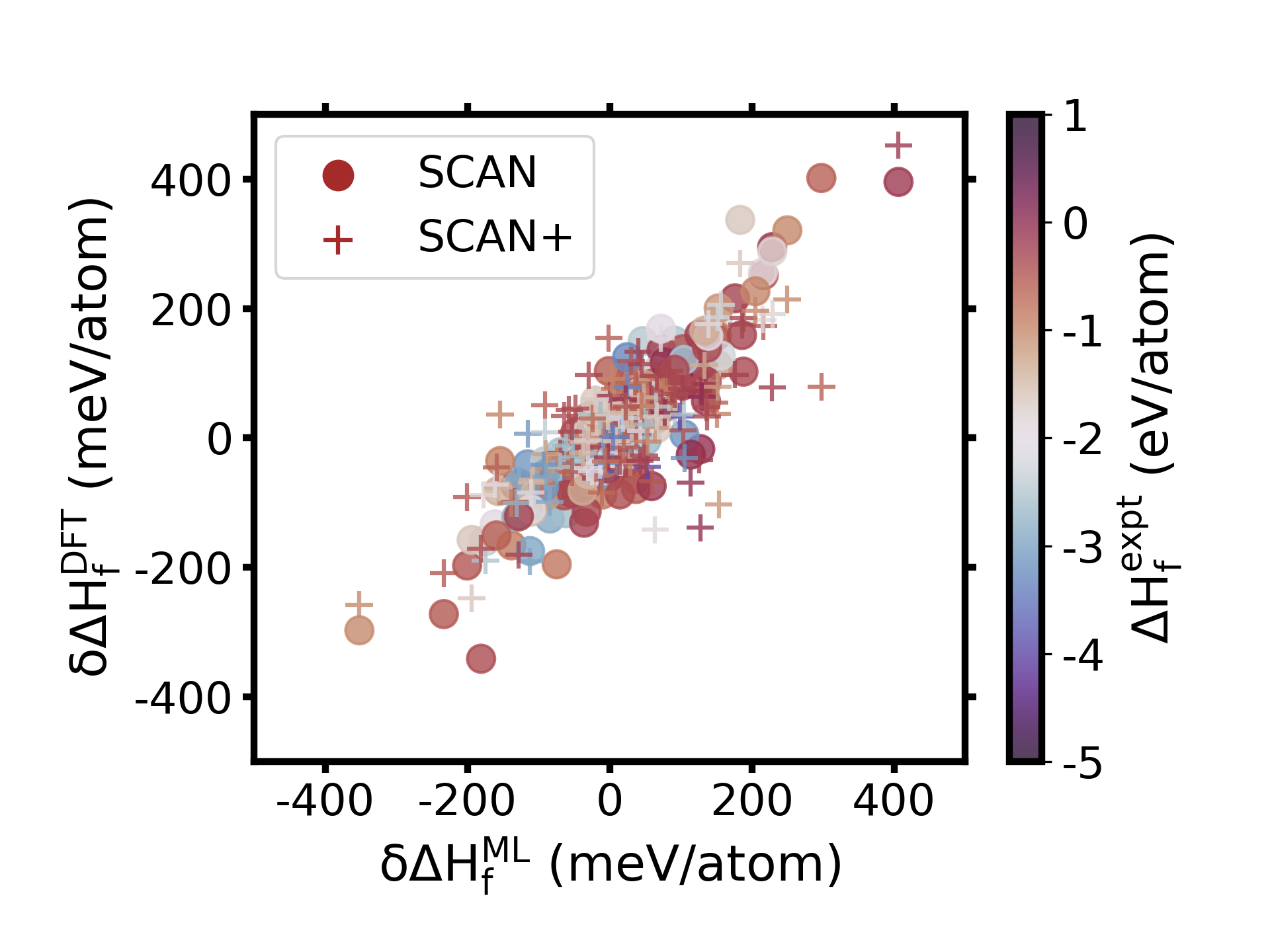}
    \end{subfigure}
    \caption{Heatmap of ${\delta}\Delta$H$_{f}^{ML}$s vs (a) ${\delta}\Delta$H$_{f}^{PBE}$s and ${\delta}\Delta$H$_{f}^{PBE+}$s and (b) ${\delta}\Delta$H$_{f}^{SCAN}$s and ${\delta}\Delta$H$_{f}^{SCAN+}$s on the corresponding 20\% left-out test set. The color used in the heatmap is based on the $\Delta$H$_{f}^{expt}$ values (in eV) as shown in the colorbar.}
    \label{pbe+ vs lap}
\end{figure}

Elemental corrections to PBE, so-called PBE+, as fit by Bartel \textit{et al.} \cite{bartel2019role} for the same set of experimental compounds have an MAE = 103 meV/atom, which is approximately 50\% lower than PBE (MAE = 195 meV/atom). Moreover, the MAEs in $\Delta$H$_{f}^{PBE+}$ are even for all binary (103, meV/atom), ternary (106 meV/atom), and quaternary (82 meV/atom) compounds, which indicates that the elemental reference energy corrections address most of the systematic errors present with PBE. However, the errors are still higher with PBE+ than with the $\Delta$-learned ML models as can be observed from Fig. \ref{pbe+ vs lap}(a) which shows a parity plot of ${\delta}\Delta$H$_{f}^{ML}$ and ${\delta}\Delta$H$_{f}^{PBE/PBE+}$ indicating that the ML model outperforms both PBE and PBE+ for the prediction of materials' $\Delta$H$_f$s. The overall reduced errors in $\Delta$H$_{f}^{ML}$ suggest that the feature set is contributing additional, material-specific knowledge, which aids in the increases the model's accuracy. \par
Moreover, as mentioned in the Introduction, databases such as OQMD, MP, and, AFLOW often use +U corrections for specific elements to mitigate the effect of the SIE and where generally ${\delta}\Delta$H$_{f}^{PBE}$ is larger. Analysis of the MAEs of PBE, PBE+, and ML (Table \ref{plus u}) shows that, PBE is most problematic for oxides of few transition metals and actinides where OQMD would use +U (365 meV/atom), followed by oxides and fluorides of transition metals (348 meV/atom) where MP would use +U. The errors are much higher for ternary and quaternary compounds (432 meV/atom (OQMD), and 418 meV/atom (MP)) compared to binary compounds. Across each category in Table \ref{plus u}, ML models show consistently lower MAE of $\sim$ 20 meV/atom compared with PBE+.
\begin{table}[h!]
\setlength{\tabcolsep}{5.0 pt}
\caption{MAE (in meV/atom) of PBE, PBE+, and ML, for the compounds in the test set (203) where OQMD, MP, and AFLOW databases would (Y) and would not (N) use GGA+U. The number inside the parenthesis represents the count of compounds.}
\resizebox{\columnwidth}{!}{%
 \begin{tabular}{lcccccc}
 \toprule
     & \multicolumn{2}{c}{(+U) OQMD} & \multicolumn{2}{c}{(+U) MP} & \multicolumn{2}{c}{(+U) AFLOW} \\
     \cmidrule(lr){2-3} \cmidrule(lr){4-5} \cmidrule(lr){6-7}
     & Y (14)        & N (189)        & Y (17)       & N (186)       & Y (98)         & N (105)        \\
     \midrule
PBE  & 365           & 182            & 348          & 181           & 198            & 191            \\
PBE+ & 213           & 95             & 209          & 94            & 124            & 84             \\
ML   & 200           & 71             & 178          & 71            & 107            & 55            \\    
                          \bottomrule
\end{tabular}}
\label{plus u}
\end{table}
\iffalse
\begin{table}[h!]
\setlength{\tabcolsep}{5.0 pt}
\caption{MAE (in meV/atom) of PBE, PBE+, and ML, for the compounds in the test set (203) where OQMD, MP, and AFLOW databases would (Y) and would not (N) use GGA+U. The number inside the parenthesis represents the count of compounds.}
\resizebox{\columnwidth}{!}{%
 \begin{tabular}{lcccccc}
 \toprule
     & \multicolumn{2}{c}{(+U) OQMD} & \multicolumn{2}{c}{(+U) MP} & \multicolumn{2}{c}{(+U) AFLOW} \\
     \cmidrule(lr){2-3} \cmidrule(lr){4-5} \cmidrule(lr){6-7}
     & Y (14)        & N (189)        & Y (37)       & N (166)       & Y (98)         & N (105)        \\
     \midrule
PBE  & 365           & 182            & 264          & 179           & 198            & 191            \\
PBE+ & 213           & 95             & 159          & 91            & 124            & 84             \\
ML   & 200           & 71             & 138          & 67            & 107            & 55            \\    
                          \bottomrule
\end{tabular}}
\label{plus u}
\end{table}
\fi
        \begin{figure}
         \centering
         \includegraphics[scale=0.355]{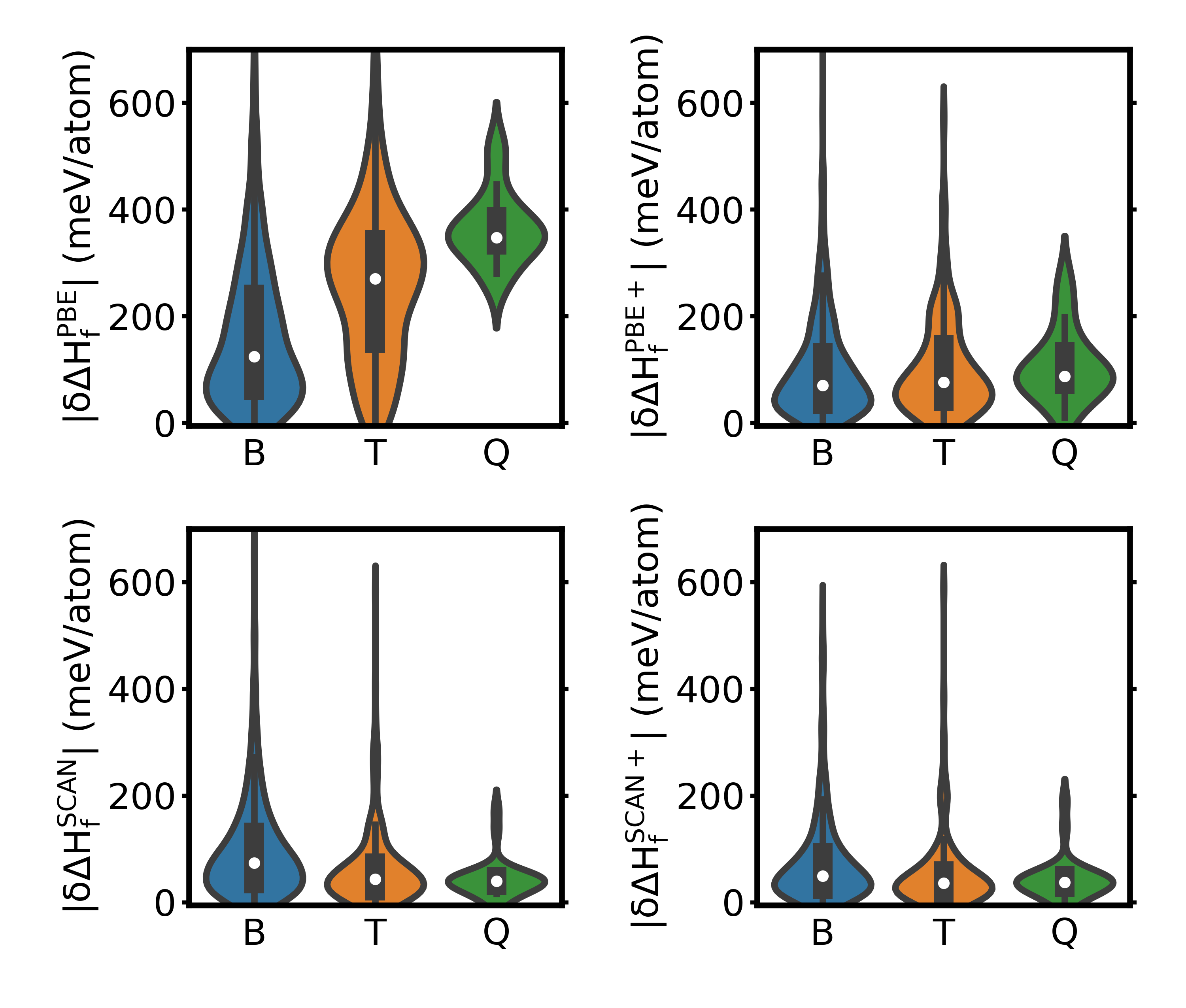} 
         \caption{Violin plots showing the distribution of absolute values of ${\delta}\Delta$H$_{f}^{PBE}$, ${\delta}\Delta$H$_{f}^{PBE+}$, ${\delta}\Delta$H$_{f}^{SCAN}$, and ${\delta}\Delta$H$_{f}^{SCAN+}$ (in meV/atom) across binary (B), ternary (T), and quaternary (Q) compositions in our dataset. Out of 1012 compounds, our dataset has 714 binary, 270 ternary, and 28 quaternary compositions.}
         \label{binary ternary}
     \end{figure} 
     
For SCAN, ${\delta}\Delta$H$_{f}^{SCAN}$s has a test-set MAE of 76 meV/atom, which is comparable to the MAE of SCAN and SCAN fitted with elemental corrections (SCAN+) of 85 and 65 meV/atom. The mean of ${\delta}\Delta$H$_{f}^{SCAN}$ is equal to zero indicating that it has no systematic bias. In contrast to what was observed for $\Delta$H$_{f}^{PBE}$, $\Delta$H$_{f}^{SCAN}$s are relatively more accurate for the ternary (MAE = 59 meV/atom) and quaternary compounds (MAE = 45 meV/atom), but less so for the binary compounds (MAE = 99 meV/atom), see Fig \ref{binary ternary}. This supports the findings in the literature that SCAN is not as accurate as PBE for intermetallics (which are primarily binary compounds in this dataset) \cite{isaacs2018performance}.\par

In order to gain an understanding of the ability of ML to substantially increase in the accuracy of the predictions of $\Delta$H$_{f}^{PBE}$, while providing little improvement for $\Delta$H$_{f}^{SCAN}$, we have trained an interpretable GAM using the same $\Delta$-learned procedure previously described for the RFR model. Despite the fact that the $\Delta$-learned ML model within the GAM framework exhibits a slightly higher error for $\Delta$H$_f$s compared to the RFR model (91 vs. 80 for SCAN; 86 meV/atom vs. 76 meV/atom for PBE, respectively, see Table \ref{perf}), the advantage of GAMs is the simple, additive structure which facilitates a more straightforward analysis of the feature set through partial dependence plots (PDP). The interpretability of GAMs is preferred over RFR, which does not consider the effect of each feature separately but instead by design incorporate interactions between features. As a result, the PDPs generated from RFR-based models may potentially show unreliable trends, and therefore, are avoided in this work.\par

\begin{figure}
    \flushleft{(a)},
    \begin{subfigure}[b]{0.50\textwidth}
         \includegraphics[scale=0.53]{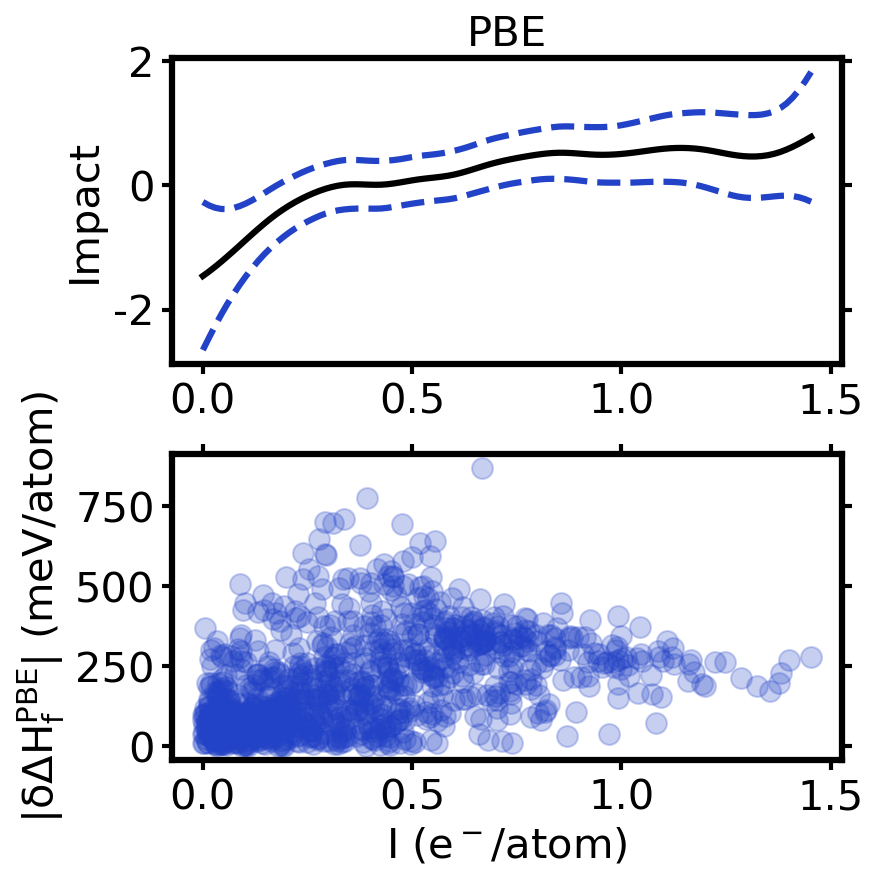}
     \end{subfigure}
     \hfill
             \flushleft{(b)}
         \begin{subfigure}[b]{0.50\textwidth}
         \includegraphics[scale=0.53]{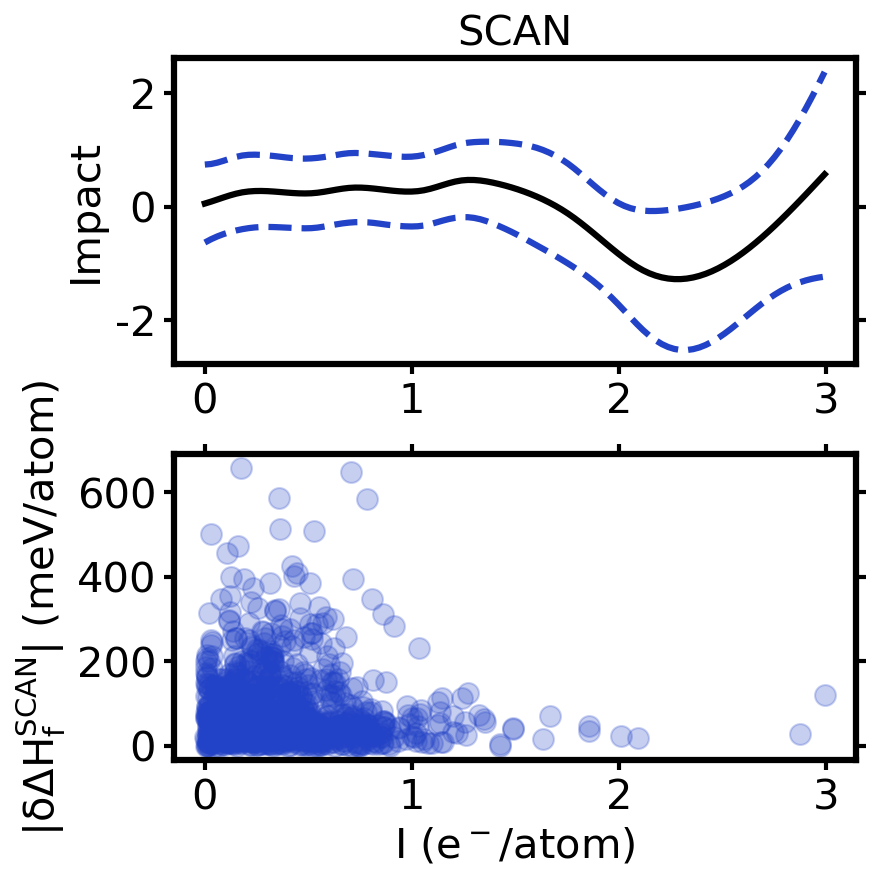} 
     \end{subfigure}         
    \caption{The impact of $I$ with units e$^{-}$/atom (solid black line) computed using the DDEC6 analysis on (a) |${\delta}\Delta$H$_{f}^{PBE}$| and (b) |${\delta}\Delta$H$_{f}^{SCAN}$|. The dotted blue line represents the 95\% confidence level. The scatter plot in the bottom panel of each subfigure displays the relation between |${\delta}\Delta$H$_{f}^{PBE/SCAN}$| (y-axis) and $I$ (x-axis).}
     \label{pdp: ionicity}
\end{figure}

Using the GAM model trained to the set of |${\delta}\Delta$H$_{f}^{PBE}$| and |${\delta}\Delta$H$_{f}^{SCAN}$| values, PDPs were generated (see Figs S26-S50) for all 25 features inputted into our model (see Table \ref{input features}). Out of those 25 different input features, PDP analysis shows that the features \textit{$\zeta$$_B$} (in e$^-$/atom), $Z$ and \textit{I} (in e$^-$/atom) have the highest impact on |${\delta}\Delta$H$_{f}^{PBE/SCAN}$|. Although \textit{$\zeta$$_B$, Z} PDPs displayed qualitatively similar trends for |${\delta}\Delta$H$_{f}^{PBE}$| and |${\delta}\Delta$H$_{f}^{SCAN}$| values, the PDP of \textit{I} indicates a stark difference. Thus, in this text, we focus our discussion on the PDP for \textit{I} (see Fig \ref{pdp: ionicity}). 

\begin{figure*}

    \captionsetup[subfigure]{labelformat=empty}
    \begin{subfigure}[b]{0.61\textwidth} \flushleft{(a)$~~~~~~~~~~~~~~~~~~~~~~~~~~~~~~~~~~~~~~~~~~~~~~~~~~~~~~~~~~~~~~~~~~~~~~~~~~~~~~~~~~~~~~~~~~~~~~~~~~~~~~~~~~~~~~~~~~~~~~~~~~~~~~~~~~~~~~~~~~~~~~~~~~~~~~~~~$(b)}
                \includegraphics[scale=0.37]{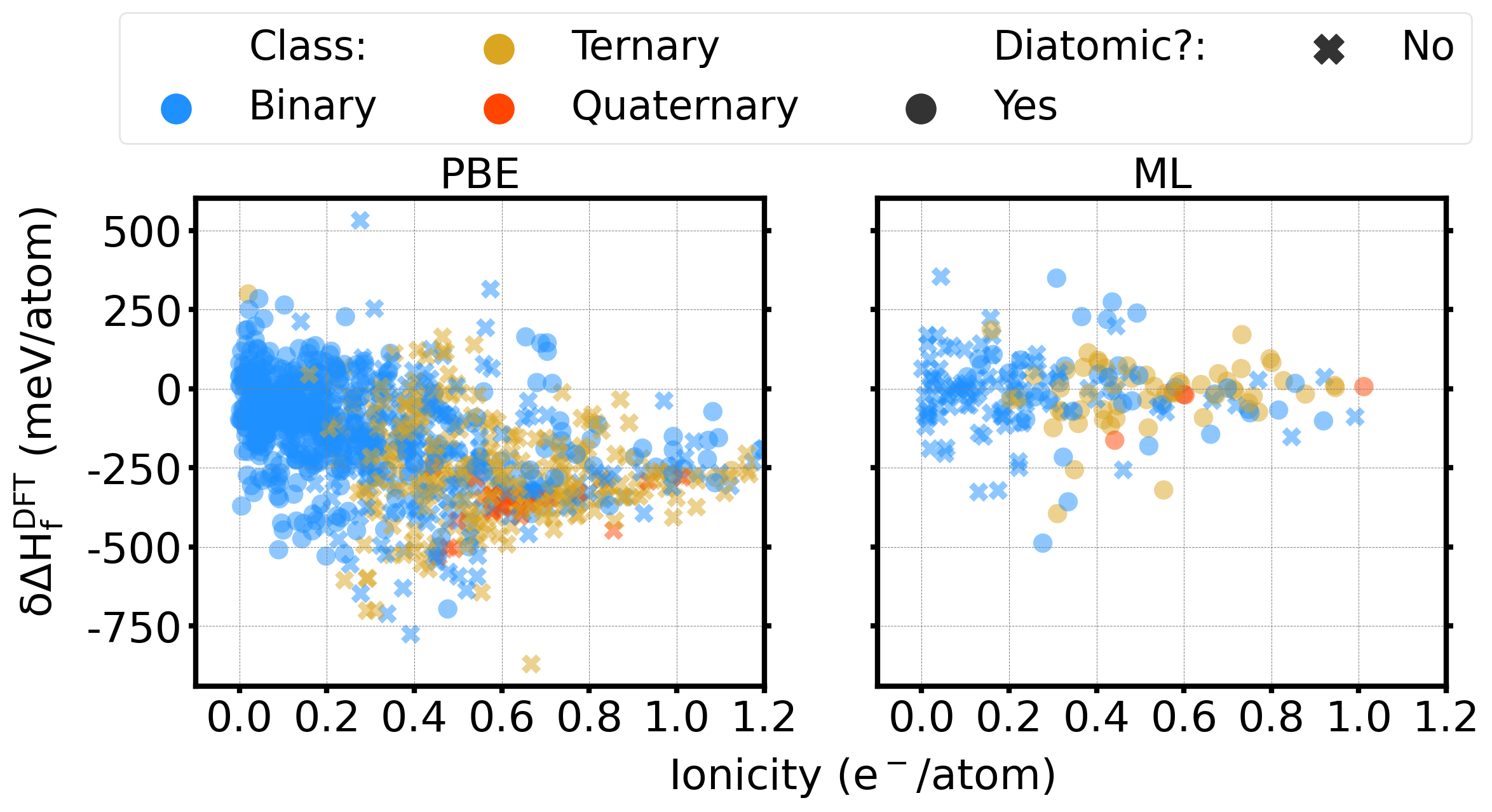}
    \end{subfigure}
        \begin{subfigure}[b]{0.38\textwidth}
                \includegraphics[scale=0.39]{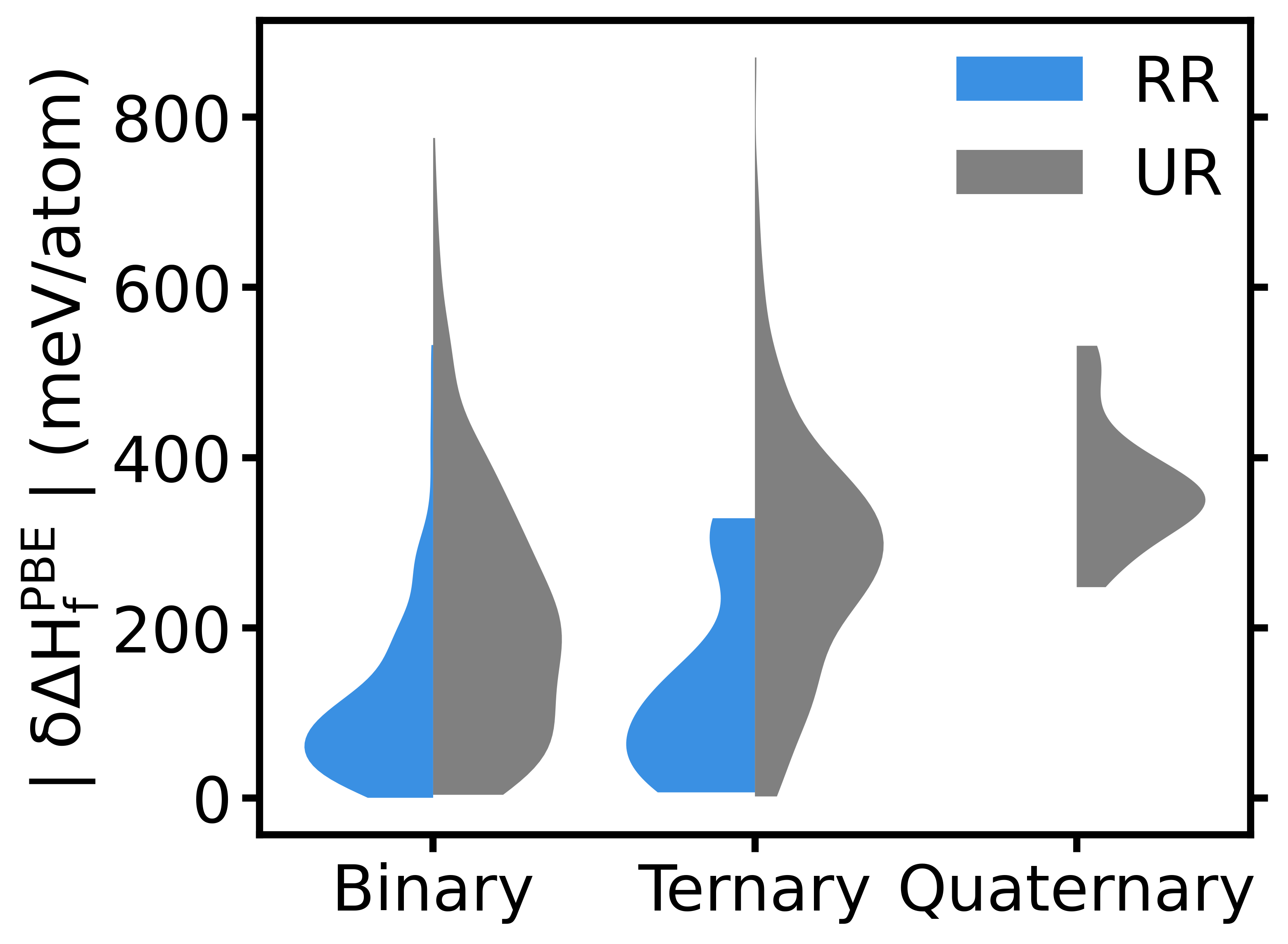}
    \end{subfigure}
    \caption{(a) shows the relation of the $I$ vs ${\delta}\Delta$H$_{f}^{PBE}$ (left; across whole dataset) and ${\delta}\Delta$H$_{f}^{ML}$ (right; across the test set). Blue, yellow, and red dots, respectively, indicate the errors for binary, ternary, and quaternary compounds. Similarly, (b) shows ${\delta}\Delta$H$_{f}^{PBE}$ as a function of binary, ternary, and quaternary compositions where PBE is reliable (RR, blue) vs. unreliable (UR, gray). The sub-group with $\zeta$$_D$<=0.43 e$^-$/atom, $P$$_D$<=0.02 e$^-$/{\AA}$^2$, and \textit{I}<=0.52 e$^-$/atom (Table \ref{input features}) is reliable (RR), the rest is unreliable (UR).
    }
    \label{reliability}
\end{figure*}

PDP plot of $I$ shows a positive impact for \textit{I} > 0.22 on |${\delta}\Delta$H$_{f}^{PBE}$|, which indicates a higher error in the calculation of $\Delta$H$_{f}^{PBE}$ for compounds with more ionic character (see Fig \ref{pdp: ionicity}). 

There are 636 compounds that have an \textit{I} > 0.22. Of which, nearly all (301 out of 314 compounds in the overall dataset) of the compounds that contain either alkali or alkaline earth elements are included in this subset. These 301 compounds have larger ${\delta}\Delta$H$_{f}^{PBE}$ values (MAE of 261 meV/atom), which is comparable to the MAE (246 meV/atom) of all 636 compounds. The other 13 of 314 alkali or alkaline earth elements compounds, which have an \textit{I} < 0.22, the MAE = 71 meV/atom for ${\delta}\Delta$H$_{f}^{PBE}$. 

Although the current literature indicates compounds containing diatomic elements (O$_2$, Cl$_2$, N$_2$, F$_2$, H$_2$) have a very high error (${\delta}\Delta$H$_{f}^{PBE}$), for the 41 (out of a total of 528) compounds within \textit{I} < 0.22, the MAE is relatively low (116 meV/atom) despite containing a few transition metal chlorides (NiCl$_2$ (390 meV/atom), TiCl$_2$ (360 meV/atom), FeCl$_3$ (300 meV/atom), FeCl$_2$ (350 meV/atom)), in this subset. In comparison, the 487/528 compounds that contain diatomic elements and are in the range \textit{I} > 0.22 have an MAE=250 meV/atom.

A similar trend is observed for binary compounds with metallic bonding (intermetallics), such as CaMg$_2$, AlNi, AlTi, where 13/25 with \textit{I} < 0.22, have an MAE of 61 meV/atom. In comparison, the remaining 12 intermetallics with an \textit{I} > 0.22, which include Ba$_2$Pb, Ca$_2$Sn, FeTi, etc.,  have significantly higher MAE (233 meV/atom). 

These results indicate that the evolution of $\Delta$H$_{f}^{PBE}$ errors and hence the corrections required for such compounds can vary based on the bonding environments. This could potentially be influenced by the number of elements (e.g., binary, ternary, or quaternary) present in the compound, which could lead to more variation in the bonding of some material.

In contrast, for |${\delta}\Delta$H$_{f}^{SCAN}$|, a highly uncertain trend with \textit{I} is observed as indicated by the flat black line in Figure \ref{pdp: ionicity}). The explicit values of the \textit{I} are also provided as subfigures in the bottom panels confirming that the qualitative trend with \textit{I} changes between the PBE and SCAN. This observation is in agreement with the previous literature that suggests SCAN improves thermochemical calculations compared with PBE by better treatment of diversely bonded systems \cite{sun2016accurate}. 

An examination of the ${\delta}\Delta$H$_{f}^{ML}$ as a function of $I$, compound class (binary, ternary, quaternary), and the presence of diatomic elements within the compound, indicates it is exactly these compounds (e.g., compounds with diatomic elements with \textit{I}>0.22) where a considerable decrease in ${\delta}\Delta$H${f}^{ML}$ compared to ${\delta}\Delta$H${f}^{PBE}$ is observed (see Fig \ref{reliability} (a)). This observation indicates there exists a specific domain formed by conditions on various properties where ${\delta}\Delta$H$_{f}^{PBE}$ is relatively low, thus constituting a reliable region. To identify such a reliable region, we employ the approach introduced in the Ref. \cite{sutton2020identifying} and identified a set of 368 compounds (out of 1011) with an MAE = 107 meV/atom for $\Delta$H$_{f}^{PBE}$ which are selected by this combination of properties: \textit{I}<=0.52, $\zeta$$_D$<=0.43, and P$_D$<=0.02 (Fig \ref{reliability}(b)). In comparison, the remaining 643 compounds have an MAE = 247 meV/atom. The constraint \textit{I}<=0.52 identified by SGD is in agreement with the analysis from the PDPs ($I$<0.22) discussed above, but expands the range of compounds in this reliable region.

Finally, because a material competes with all possible phases in its compositional space (enthalpy of decomposition; $\Delta$H$_{d}$) for stability rather than just the elemental phases ($\Delta$H$_{f}$),\cite{bartel2019role,bartel2020critical} we assess how the lower errors in ${\delta}\Delta$H$_{f}^{ML}$ translates to $\Delta$H$_d$s (i.e., $\Delta$H$_{d}^{ML}$) using the \textit{leave-one-chemical-space-out} scheme and compared them with the $\Delta$H$_d$s obtained using $\Delta$H$_{f}^{expt}$ ($\Delta$H$_{d}^{expt}$) and $\Delta$H$_{f}^{DFT}$ ($\Delta$H$_{d}^{DFT}$). 

As analyzed in Ref \cite{bartel2019role}, the majority of compounds in the MP database \cite{jain2013commentary} compete for stability with compound phases (Type 2) or a mixture of compounds and elemental phases (Type 3) rather than elemental phases only (Type 1). Although $\Delta$H$_{d}^{DFT}$s (Type 2 and Type 3) are typically 1-2 orders of magnitude smaller compared to $\Delta$H$_{f}^{DFT}$s, previous literature suggests that the ML models trained to predict accurate $\Delta$H$_{f}^{DFT}$ values perform poorly for $\Delta$H$_{d}^{DFT}$s. \cite{bartel2020critical} This is because errors made by ML models for $\Delta$H$_{f}^{DFT}$ are not as systematic as the DFT errors with respect to the chemical composition \cite{bartel2020critical,wang2021framework}. In contrast, our results show that (Table S1) the $\Delta$H$_{d}^{ML}$ error is comparable to $\Delta$H$_{d}^{PBE}$ error for both type 2 and type 3 compounds suggesting that ML corrections to $\Delta$H$_{f}^{PBE}$ are mostly canceling out for stability prediction ($\Delta$H$_{d}$s).  We observe similar findings for $\Delta$H$_{d}^{ML}$s in the case of SCAN too (Table S2). 

\section*{Conclusion}
In this work, we used ML to correct the DFT-calculated enthalpy of formation ($\Delta$H$_{f}^{DFT}$) to predict experimental enthalpy of formation ($\Delta$H$_{f}^{expt}$) of 1011 solids. The ML model inputs 25 numerical properties that capture electronic structure information computed with either the PBE or SCAN density functional. Among various ML methods, random forest regressor (RFR) performs best and corrects $\Delta$H$_{f}^{PBE}$ to within an MAE = 80 meV/atom of $\Delta$H$_{f}^{expt}$, significantly improving upon the MAE of 195 meV/atom of $\Delta$H$_{f}^{PBE}$s. In contrast, the ML model correcting $\Delta$H$_{f}^{SCAN}$ (MAE=76 meV/atom) has comparable accuracy to SCAN (85 meV/atom).

To analyze why a significant correction is observed for PBE while not so for SCAN, we studied the impact and trends of all 25 features on the absolute difference between $\Delta$H$_{f}^{expt}$ and $\Delta$H$_{f}^{DFT}$ (errors), using the partial dependence plots (PDPs) within generalized additive models. PDPs showed that for a number of features, there is a clear trend between feature values and the $\Delta$H$_{f}^{PBE}$ errors indicating that the errors made by PBE are more systematic and straightforward to correct.  More specifically, the interpretable PDP analysis identified the ionicity (I), which quantifies the ratio of average charge transferred and summed bond orders, as a strong indicator for high $\Delta$H$_{f}^{PBE}$ errors. Based on the PDP analysis, we explicitly showed that the MAE for compounds with \textit{I} < 0.22 (113 meV/atom) is a factor 2 smaller compared to the ones with \textit{I} > 0.22 (246 meV/atom) indicating different evolution of errors across different regions. This trend from the PDPs is further supported by the application of subgroup discovery (SGD) to the PBE errors using the same set of 25 features to identify the reliable regions (i.e., lowest errors). Out of these 25 features, SGD selects a subgroup with the lowest error (108 meV/atom), containing 368 compounds out of 1011 total, and is predominantly comprised of compounds containing low charge transfer between atoms based on the selector: \textit{I}<=0.52, $\zeta_{D}$<=0.43, and $P_D$<=0.02, containing about one-third of compounds (368/1011) with $\Delta$H$_{f}^{PBE}$ errors less than 108 meV/atom. Here $\zeta$ and $P$ quantify the average charge transfer and dipole moment per volume from DDEC6 analysis.  Interestingly, although the literature suggests PBE is reliable for intermetallics and less so for oxides and halides, our analysis reveals intermetallics pose a challenge for PBE only when the charge transfer is significant (I>0.22) and oxides and halides may be more amenable to PBE if the charge transfer is relatively low (I<0.22), for example, compounds such as MoCl$_5$ (98 meV/atom), PtCl$_3$ (2 meV/atom), Ag$_2$O (29 meV/atom), PdO (116 meV/atom), etc. Overall, our work leads to a better understanding of the strengths and limitations of the semilocal density functional predictions of materials thermochemistry and can potentially guide the future development of more accurate methods.

\begin{acknowledgments}
We acknowledge support by departmental start-up funds at the University of South Carolina and the NSF-funded MadeinSC (Award number OIA-1655740).
\clearpage
\end{acknowledgments}

\bibliography{main.bib}% Produces the bibliography via BibTeX.

\end{document}